\documentclass[a4paper,11pt]{article}

\usepackage{contribution}
\newcommand{\weblink}[2][]{%
    \ifthenelse{\equal{#1}{}}%
    {\textnormal{\url{#2}}}%
    {\textnormal{\href{#2}{#1}}}%
}

\newcommand{\acknowledgements}[1]{%
  \bigskip\bigskip
  \textsf{\textbf{\Large Acknowledgements}} \\[2ex]
  {#1}
  \bigskip
}

\def\beq{\begin{equation}}
\def\eeq#1{\label{#1}\end{equation}}
\def\eeqn{\end{equation}}

\def\beqa{\begin{eqnarray}}
\def\eeqa#1{\label{#1}\end{eqnarray}}
\def\eeqan{\end{eqnarray}}

\let\bar=\overbar

\def\Dslash{\not{\hbox{\kern-4pt $D$}}}
\def\dslash{\not{\hbox{\kern-2pt $\del$}}}

\def\msb{{\bar{\ssstyle M \kern -1pt S}}}

\newcommand{\contribution}[7][]{%
  \clearpage
  \thispagestyle{plain}
  \ifthenelse{\equal{#1}{}}
  {\hypersetup{pdftitle={#2}}}
  {\hypersetup{pdftitle={#1}}}
  \hypersetup{pdfauthor={{#3} {#4}}}
  {\centering\normalfont\LARGE\bfseries\sffamily #2 \par\nobreak}
  \lhead{}
  \chead{%
    \textit{\footnotesize XIV International Conference on Hadron Spectroscopy
      (\weblink[\textit{hadron2011}]{http://www.hadron2011.de}), 13-17 June 2011, Munich, Germany}%
  }
  \rhead{}
  \bigskip
  \begin{center}
    {#3} {#4}\ifthenelse{\equal{#6}{}}{}{\footnote{\weblink[#6]{mailto:#6}}}
    \ifthenelse{\equal{#7}{}}{}{#7} \\
    \textit{#5}
  \end{center}
  \bigskip
}

\renewcommand{\abstract}[1]{%
  \begin{center}
    \begin{minipage}{0.85\textwidth}
      \begin{footnotesize}
        #1
      \end{footnotesize}
    \end{minipage}
  \end{center}
  \bigskip
}

\begin{document}

% % % % % % % % % % % % % % % % % % % % % % % % % % % % % % % % % % % % % % % % %
% your proceedings
%%%%%%%%%%%%%%%%%%%%%%%%%%%%%%%%%%%%%%%%%%%%%%%%%%%%%%%%%%%%%%%%%%%%%%%%%%%%%%%%%
%
% template for hadron2011 contribution
%
% please do not rename this file
%
% to create document run
%
%     pdflatex hadron2011.tex
%
%%%%%%%%%%%%%%%%%%%%%%%%%%%%%%%%%%%%%%%%%%%%%%%%%%%%%%%%%%%%%%%%%%%%%%%%%%%%%%%%%
{  % do not remove

%%%%%%%%%%%%%%%%%%%%%%%%%%%%%%%%%%%%%%%%%%%%%%%%%%%%%%%%%%%%%%%%%%%%%%%%%%%%%%%%%
% please define your macros here

%
%%%%%%%%%%%%%%%%%%%%%%%%%%%%%%%%%%%%%%%%%%%%%%%%%%%%%%%%%%%%%%%%%%%%%%%%%%%%%%%%%

%\preprint{\hbox{TUM-EFT 24/11}}
%%%%%%%%%%%%%%%%%%%%%%%%%%%%%%%%%%%%%%%%%%%%%%%%%%%%%%%%%%%%%%%%%%%%%%%%%%%%%%%%%
% define title, author, and address
% contribution[short title]{title}{author first name}{author last name}{author address}{author email}{collaboration}
% the short title will appear in the page headers and the TOC of the book of proceedings
% the last two arguments may be left empty
\contribution  % short title (optional)
{\hfill {\tt \small TUM-EFT 24/11} \\Electric dipole transitions of heavy quarkonium in pNRQCD}
{Piotr}{Pietrulewicz}  % first and last name of author
{Physik-Department, Technische Universit\"at M\"unchen \\
 James-Franck-Str. 1, 85748 Garching, Germany}  % author address
{}  % author email optional
{}

%%%%%%%%%%%%%%%%%%%%%%%%%%%%%%%%%%%%%%%%%%%%%%%%%%%%%%%%%%%%%%%%%%%%%%%%%%%%%%%%%

%%%%%%%%%%%%%%%%%%%%%%%%%%%%%%%%%%%%%%%%%%%%%%%%%%%%%%%%%%%%%%%%%%%%%%%%%%%%%%%%%
% abstract
\abstract{%
We propose a systematic, model-independent treatment of electric dipole transitions of heavy quarkonium. Within  an effective field theory, concretely potential non-relativistic QCD, the relativistic corrections of relative order $v^2$ to the decay rate are derived. An existing formalism developed for M1 decays will be extended for our purpose. We scrutinize and complement former results from potential model calculations.
}
%
%%%%%%%%%%%%%%%%%%%%%%%%%%%%%%%%%%%%%%%%%%%%%%%%%%%%%%%%%%%%%%%%%%%%%%%%%%%%%%%%%

%%%%%%%%%%%%%%%%%%%%%%%%%%%%%%%%%%%%%%%%%%%%%%%%%%%%%%%%%%%%%%%%%%%%%%%%%%%%%%%%%
% main text
% for short contributions sections are optional
\section{Introduction}

Radiative transitions play an important role for our understanding of QCD, in particular of heavy quarkonium. They provide information about the wave functions describing the physical system and probe both the perturbative and non-perturbative regime. Especially E1 transitions give significant contributions to the total decay rate and are observed in the experimental facilities. Recently, decays of charmonium were measured at BES and CLEO, including the observation of the process $h_c \rightarrow \eta _c \, \gamma$ in 2010 \cite{h_c}. Concerning bottomonium, CLEO, BaBar and Belle produced many data, e.g. for the determination of the branching fractions $\chi_b$ states \cite{chi_b, Babar_bottomonium}. A review about recent developments can be found in \cite{QWG}.

On the theory side, electric dipole transitions were treated in several potential models, a summary can be found in \cite{Eichten_Godfrey}. We will refer to \cite{Grotch} for comparison with the general results for $\chi$-decays. A model-independent treatment to check and improve the calculations has been missing so far. However, in the last decade there has been significant progress using effective field theories (EFTs) to describe heavy quarkonium (see \cite{quarkonium_review} and references therein). Since heavy quarkonium is assumed to be a non-relativistic system we may take advantage of the hierarchy of scales $m \gg mv \gg mv^2$, where $v \ll 1$ is the heavy quark velocity, $m$ is the heavy quark mass ("hard scale"), $p \sim mv$ is the relative momentum of the bound state ("soft scale") $E \sim mv^2$ is the binding energy $E \sim mv^2$ ("ultrasoft scale"). The ultimate EFT living at the ultrasoft scale is potential non-relativistic QCD (pNRQCD). In 2005, for the first time radiative decays, concretely M1 transitions, were calculated in this theory \cite{M1_Brambilla}. Using the framework of that paper as a guideline we close the gap and compute the decay rates of the E1 processes $n^3 P_J \rightarrow {n'}^3 S_1 \, \gamma$ and $n^1 P_1 \rightarrow {n'}^1 S_0 \, \gamma$. The following is based on \cite{E1_paper}.

\section{The Lagrangians in NRQCD and pNRQCD}

By integrating out the hard scale $m \gg \Lambda_{QCD}$ from the fundamental theory, QCD, in perturbation theory ($\alpha_s (m) \ll 1$) one obtains non-relativistic QCD (NRQCD) \cite{NRQCD,Bodwin}. For the calculation of E1 transitions at NLO only the two-fermion Lagrangian $\mathcal{L}_{2-f}$ matters and the relevant part reads 

\begin{equation}
\mathcal{L}_{2-f} = \psi^{\dagger} \left( iD_{0} + \frac{{\bf D}^2}{2m} + \frac{{\bf D}^4}{8m^3} \right) \psi + e e_Q  \psi^{\dagger} \left( \frac{c_F^{em}}{2m} {\bf \boldsymbol \sigma \cdot B}^{em} +i \frac{c_s^{em}}{8m^2} {\bf \boldsymbol \sigma \cdot } [{\bf D} \times,{\bf  E}^{em}] \right) \psi + c.c. \, .
\end{equation}
with $iD_0 = i\partial _0 - g T^a A_0^a - e e_Q A_0^{em}$, $i{\bf D} = i\boldsymbol \nabla + g T^a {\bf A}^a + e e_Q {\bf A}^{em}$ and $\psi$ denoting a Pauli spinor for the heavy quark. The matching coefficients are found to be $c_F^{em} = 1 + C_F \frac{\alpha_s}{2 \pi} + \mathcal{O}(\alpha_s^2) \label{cF}$, $c_s^{em} = 2 c_F^{em} -1 \label{cs}$. 

For processes at the ultrasoft scale, NRQCD is not yet the appropriate theory, since there are still several scales entangled ($p, E, \Lambda_{QCD}$) and thus no homogeneous power counting can be established. Integrating out the soft scale $mv$ we obtain a theory for ultrasoft modes, i.e. pNRQCD \cite{pNRQCD:Pineda,pNRQCD:Brambilla}. The crucial step to disentangle the energy and momentum scale is the multipole expansion in the relative distance $r$. To be definite we will work in the weak-coupling regime, where $p \gg E \gtrsim \Lambda_{QCD}$. The power counting reads
\begin{equation}
r \sim 1/mv, \, \boldsymbol \nabla _r \equiv \partial /\partial {\bf r} \sim mv, \, \boldsymbol \nabla \equiv \partial /\partial {\bf R} \sim mv^2 \, , \,{\bf E},  {\bf B} \sim (mv^2)^2, \, {\bf E}^{em} , {\bf B}^{em} \sim k_{\gamma}^2 \label{weak}\, .
\end{equation}
$k_{\gamma}$ is the energy of the emitted photon, which scales like $k_{\gamma} \sim mv^2$ for transitions between states with different principal quantum numbers.

The pNRQCD-Lagrangian contributing at NLO in the decay rate, i.e. at order $ k^{3}_{\gamma} v ^0 / m^2$, reads 

\begin{align}\ \label{L_pNRQCD}
\mathcal{L}_{\textrm{pNRQCD}}  = & \int d^3 r \mathrm{ Tr} \left\{ S^{\dagger} \left( i {\partial}_0 + \frac{{\boldsymbol \nabla}^2}{4m} +\frac{{\boldsymbol \nabla}_r^2}{m}+ \frac{{\boldsymbol \nabla}_r^4}{4m^2} - V_S \right) S +  O^{\dagger} \left( i D_0 + \frac{{\bf D}^2}{4m} +  \frac{{\boldsymbol \nabla}_r^2}{m} - V_O \right) O \right. \nonumber\\
& \hspace{13mm} + \left. g V_A( O^{\dagger} {\bf r}\cdot {\bf E} S + S^{\dagger} {\bf r} \cdot {\bf E} O) \right\} \nonumber \\
&  + \mathcal{L}_{\gamma \textrm{pNRQCD}} + \mathcal{L}_{\textrm{light}}  \, ,
\end{align}
where the covariant derivatives are given by $iD_0 O= i\partial _0 O - g [T^a A_0^a, O]$ and $i{\bf D} O = i\boldsymbol \nabla O + g [T^a {\bf A}^a, O]$ and the trace goes over the color and spin indices. The singlet potential $V_S$ has been calculated perturbatively and non-perturbatively to order $1/m^2$ (\cite{qspectrum, pNRQCD_1m, pNRQCD_1m2}, for more original references see \cite{quarkonium_review}), we display the structure of the relevant potentials for computations at NLO in the decay rate,
\begin{align}
V_S (r) = & \, V^{(0)}(r) + \frac{V^{(1)}_r(r)}{m} + \frac{V^{(2)}_{SI}(r)}{m^2}+ \frac{V^{(2)}_{SD}(r)}{m^2} \, , \label{V1} \\
V^{(2)}_{SI}(r) = & \, V^{(2)}_r(r) + \frac{1}{2} \{ V^{(2)}_{p^2}(r), {\bf p}^2 \} + \frac{V^{(2)}_{L^2}(r)}{r^2} {\bf L}^2 \, , \label{V2} \\
V^{(2)}_{SD}(r) = & V^{(2)}_{LS}(r){\bf L}\cdot {\bf S} +V^{(2)}_{S^2}(r){\bf S}^2 + V^{(2)}_{S_{12}}(r) \left[3 (\hat{\bf r}\cdot \boldsymbol\sigma _1)(\hat{\bf r}\cdot \boldsymbol\sigma _2) - \boldsymbol\sigma _1 \cdot \boldsymbol\sigma _2 \right] \label{V3} \, .
\end{align}
The relevant part of $\mathcal{L}_{\gamma \textrm{pNRQCD}}$ for E1 transitions is 
\begin{align}\label{LagE1}
\mathcal{L}_{\gamma \textrm{pNRQCD}}^{E1} =  e e_Q \int d^3 r \, \mathrm{Tr} & \left\{ V^{r\cdot E} S^{\dagger} {\bf r}\cdot {\bf E}^{em} S + V_O ^{r\cdot E} O^{\dagger} {\bf r}\cdot {\bf E}^{em} O + \frac{1}{24} V^{(r\nabla)^2  r \cdot E}  S^{\dagger} {\bf r}\cdot [({\bf r \boldsymbol \nabla})^2  {\bf E}^{em}] S \right. \nonumber\\
& +i \frac{1}{4m}V^{\nabla \cdot (r \times B)}  S^{\dagger} \{ \boldsymbol \nabla \cdot , {\bf r} \times {\bf B}^{em} \} S \nonumber \\
&+ i \frac{1}{12m} V^{\nabla _r \cdot (r \times (r\nabla)B)}  S^{\dagger} \{ \boldsymbol \nabla _r \cdot , {\bf r} \times  [({\bf r \boldsymbol \nabla}) {\bf B}^{em}] \} S \nonumber \\
& + \frac{1}{4m} V^{(r \nabla) \sigma \cdot B} [ S^{\dagger}, \boldsymbol \sigma ] \cdot [({\bf r \boldsymbol \nabla}) {\bf B}^{em}] S \nonumber \\
& \left. -i \frac{1}{4m^2}V^{\sigma \cdot(E \times \nabla _r)} [ S^{\dagger}, \boldsymbol \sigma] \cdot ({\bf E}^{em} \times \boldsymbol \nabla _{r}) S \right\} \, . 
\end{align}
In fact more terms are allowed according to the symmetries of pNRQCD. However, we can show that their matching coefficients vanish. The matching is done by equating Green's functions in NRQCD and pNRQCD at the energy scale $mv$ order by order in the inverse mass and $r$.
\begin{figure}
  \begin{center}
  \includegraphics[width=0.9 \textwidth]{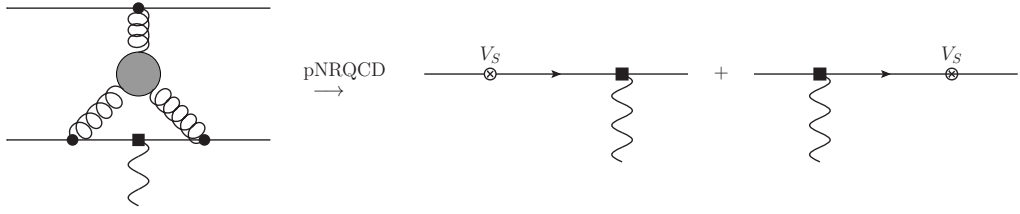}
  \caption{Example for a reducible diagram, if the electromagnetic operator commutes with the gluonic ones. It does not contribute to the matching coefficient of a single operator.}
  \end{center}
  \label{reducible}
\end{figure}
The crucial argument for several operators is that diagrams in NRQCD which can be cast into a reducible structure also give reducible diagrams in pNRQCD. Therefore they have to be subtracted to obtain irreducible operators in pNRQCD and do not play a role in the matching procedure. An example is the diagram in Fig. 1, where the gluonic contribution can be factorized out yielding just a potential.
Using this argument we can fix all of the Wilson coefficients in (\ref{LagE1}), so that the exact QCD results reproduce the ones from tree level calculations, namely
\begin{equation}
V^{r\cdot E} = V_O ^{r\cdot E} = V^{(r\nabla)^2 r \cdot E}= V^{\nabla \cdot (r \times B)}  = V^{(r\nabla)\nabla _r \cdot (r \times B)} = 1, \, V^{(r \nabla) \sigma \cdot B} = c_F^{em}, \, V^{\sigma \cdot(E \times \nabla _r)} = c_s^{em} \, .
\end{equation}

\section{Results}

With the help of the formalism developed in \cite{M1_Brambilla} we can describe the states in a quantum mechanical way using wave functions and compute the decay rate at NLO from the Lagrangian (\ref{LagE1}). We obtain
\begin{equation}\label{E1_final}
 \Gamma _{n^3 P_J \rightarrow {n'}^3 S_1 \gamma} = \frac{4}{9} \, \alpha _{em} e_Q ^2 k_{\gamma}^3 I_3 ^2(n1 \rightarrow n'0) \left( 1 + R - \frac{k_{\gamma}^2}{60} \frac {I_5}{I_3} - \frac{k_{\gamma}}{6m} + \frac{k_{\gamma} (c_F^{em}-1)}{2m} \left[\frac{J(J+1)}{2} -2 \right] \right) \, ,
\end{equation}
where
\begin{equation}
  I_N \equiv \int _0 ^{\infty} dr  \, r^N R_{n'0} (r) R _{n1} (r) \, .
\end{equation}
$R$ contains all of the wave-function corrections due to the higher-order potentials mentioned in (\ref{V1})-(\ref{V3}), the relativistic correction of the kinetic energy, $-{\bf p}^4/4m^3$, and higher-order Fock state contributions which are given by the diagrams in Fig. \ref{color-octet}. In contrast to M1 transitions the latter ones do not vanish for E1 decays.

\begin{figure}
  \begin{center}
  \includegraphics[width=0.9\textwidth]{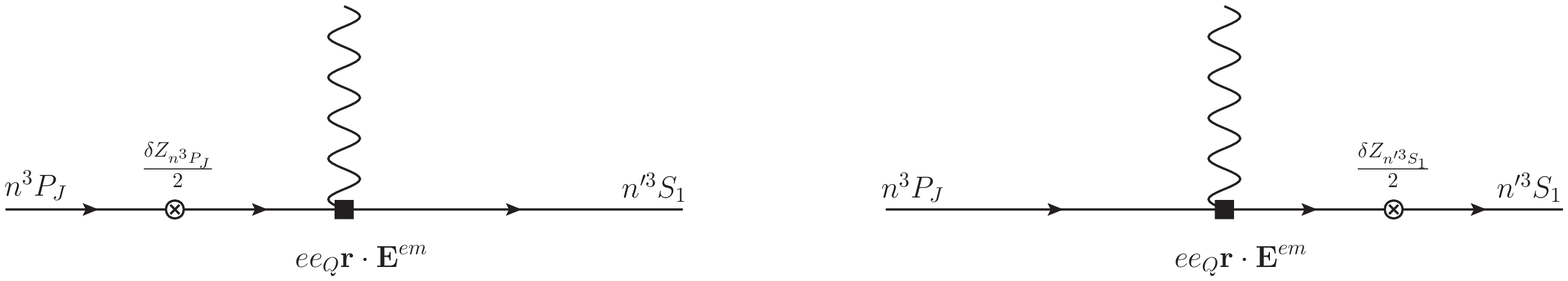}
  \includegraphics[width=0.9\textwidth]{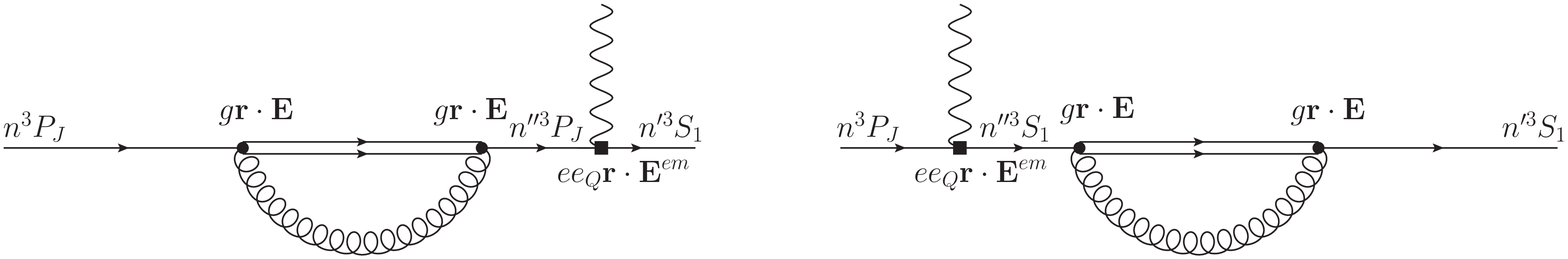}
  \includegraphics[width=0.45\textwidth]{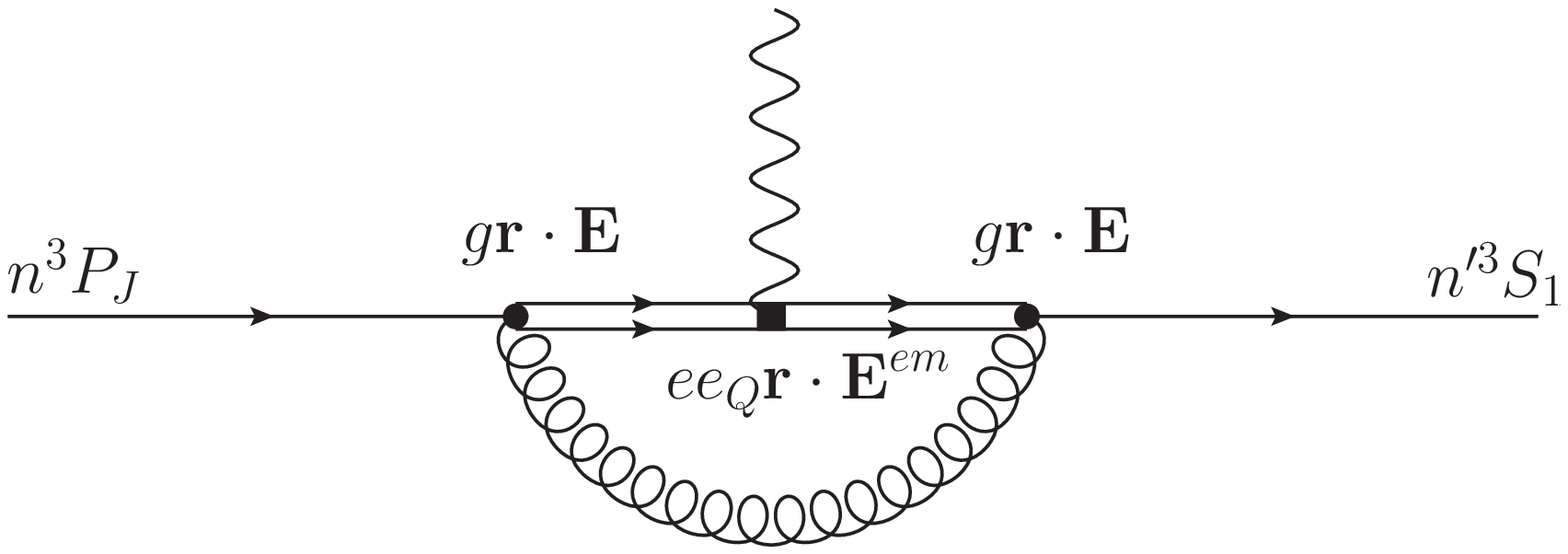}
  \caption{Color octet contributions to E1 transitions, the double line represents the intermediate octet state.}
  \label{color-octet}
  \end{center}
\end{figure}

The expression (\ref{E1_final}) is also valid in the strongly coupled regime (without color-octet contributions in $R$), where $p \sim \Lambda_{QCD}$, since we made use of non-perturbative matching arguments and additional operators do not appear in this regime.

Compared to the results with the potential model calculation in \cite{Grotch} we find an equivalence between (\ref{E1_final}) and the corresponding formula there at the given order. However, our definite power counting allowed us to include all relativistic corrections systematically, in particular the color-octet contributions in the weak-coupling regime and the one coming from $V_r ^{(1)}$. Both were missing in former approaches. Furthermore we can show that the anomalous magnetic moment $c_F^{em}-1 \sim \mathcal{O}(\alpha_s(m))$ is actually suppressed and does lead to large non-perturbative contributions.

Without much effort one can extend the discussion to other processes like $n^1 P_1 \rightarrow {n'}^1 S_0 \gamma$ and $n^3 S_1 \rightarrow {n'}^3 P_J \gamma$. Finally, based on these results a phenomenological analysis for charmonium and bottomonium decays can be performed.

\acknowledgements{
  I would like to thank Nora Brambilla and Antonio Vairo for the collaboration on this work. 
}

%%%%%%%%%%%%%%%%%%%%%%%%%%%%%%%%%%%%%%%%%%%%%%%%%%%%%%%%%%%%%%%%%%%%%%%%%%%%%%%%%
% acknowledgements (optional)

%%%%%%%%%%%%%%%%%%%%%%%%%%%%%%%%%%%%%%%%%%%%%%%%%%%%%%%%%%%%%%%%%%%%%%%%%%%%%%%%%
% bibliographic items can be constructed using the LaTeX format in SPIRES
% see http://www.slac.stanford.edu/spires/hep/latex.html
% SPIRES will also supply the CITATION line information; please include it

%
%%%%%%%%%%%%%%%%%%%%%%%%%%%%%%%%%%%%%%%%%%%%%%%%%%%%%%%%%%%%%%%%%%%%%%%%%%%%%%%%%

}  % do not remove

%%% Local Variables: 
%%% mode: latex
%%% TeX-master: "../hadron2011.tex"
%%% End: 

%\input{contribution}


\begin{thebibliography}{99}
  
\bibitem{h_c}
  M.~Ablikim {\it et al.} [ The BESIII Collaboration ],
  %``Measurements of h_c(^1P_1) in psi' Decays,''
  Phys.\ Rev.\ Lett.\  {\bf 104 } (2010)  132002.
  [arXiv:1002.0501 [hep-ex]].

\bibitem{chi_b}
  M.~Kornicer {\it et al.} [ The CLEO Collaboration ],
  %``Measurements of branching fractions for electromagnetic transitions involving the $\chi_{bJ}(1P)$ states,''
  Phys.\ Rev.\  {\bf D83 } (2011)  054003.
  [arXiv:1012.0589 [hep-ex]].
  
\bibitem{Babar_bottomonium}
  J.~P.~Lees {\it et al.} [ The BABAR Collaboration ],
  %``A Study of radiative bottomonium transitions using converted photons,''
  [arXiv:1104.5254 [hep-ex]].

\bibitem{QWG}
  N.~Brambilla, S.~Eidelman, B.~K.~Heltsley, R.~Vogt, G.~T.~Bodwin, E.~Eichten, A.~D.~Frawley, A.~B.~Meyer {\it et al.},
  %``Heavy quarkonium: progress, puzzles, and opportunities,''
  Eur.\ Phys.\ J.\  {\bf C71 } (2011)  1534.
  [arXiv:1010.5827 [hep-ph]].
 
\bibitem{Eichten_Godfrey}
  E.~Eichten, S.~Godfrey, H.~Mahlke, J.~L.~Rosner,
  %``Quarkonia and their transitions,''
  Rev.\ Mod.\ Phys.\  {\bf 80 } (2008)  1161-1193.
  [hep-ph/0701208].
 
 \bibitem{Grotch}
  H.~Grotch, D.~A.~Owen, K.~J.~Sebastian,
  %``Relativistic Corrections To Radiative Transitions And Spectra Of Quarkonia,''
  Phys.\ Rev.\  {\bf D30 } (1984)  1924.

\bibitem{quarkonium_review}
  N.~Brambilla, A.~Pineda, J.~Soto, A.~Vairo,
  %``Effective field theories for heavy quarkonium,''
  Rev.\ Mod.\ Phys.\  {\bf 77 } (2005)  1423.
  [hep-ph/0410047].

\bibitem{M1_Brambilla}
  N.~Brambilla, Y.~Jia, A.~Vairo,
  %``Model-independent study of magnetic dipole transitions in quarkonium,''
  Phys.\ Rev.\  {\bf D73 } (2006)  054005.
  [hep-ph/0512369].

\bibitem{E1_paper}
  N.~Brambilla, P.~Pietrulewicz and A.~Vairo,
  %``Model-independent Study of Electric Dipole Transitions in Quarkonium,''
  arXiv:1203.3020 [hep-ph].
  %%CITATION = ARXIV:1203.3020;%%
 
\bibitem{NRQCD}
  W.~E.~Caswell, G.~P.~Lepage,
  %``Effective Lagrangians for Bound State Problems in QED, QCD, and Other Field Theories,''
  Phys.\ Lett.\  {\bf B167 } (1986)  437.
 
\bibitem{Bodwin}
  G.~T.~Bodwin, E.~Braaten, G.~P.~Lepage,
  %``Rigorous QCD analysis of inclusive annihilation and production of heavy quarkonium,''
  Phys.\ Rev.\  {\bf D51 } (1995)  1125-1171.
  [hep-ph/9407339].

\bibitem{pNRQCD:Pineda}
  A.~Pineda, J.~Soto,
  %``Effective field theory for ultrasoft momenta in NRQCD and NRQED,''
  Nucl.\ Phys.\ Proc.\ Suppl.\  {\bf 64 } (1998)  428-432.
  [hep-ph/9707481].

\bibitem{pNRQCD:Brambilla}
  N.~Brambilla, A.~Pineda, J.~Soto, A.~Vairo,
  %``Potential NRQCD: An Effective theory for heavy quarkonium,''
  Nucl.\ Phys.\  {\bf B566 } (2000)  275.
  [hep-ph/9907240].

\bibitem{qspectrum}
  N.~Brambilla, A.~Pineda, J.~Soto, A.~Vairo,
  %``The Heavy quarkonium spectrum at order m alpha**5(s) l n alpha(s),''
  Phys.\ Lett.\  {\bf B470 } (1999)  215.
  [hep-ph/9910238].

\bibitem{pNRQCD_1m}
  N.~Brambilla, A.~Pineda, J.~Soto, A.~Vairo,
  %``The QCD potential at O(1/m),''
  Phys.\ Rev.\  {\bf D63 } (2001)  014023.
  [hep-ph/0002250].

\bibitem{pNRQCD_1m2}
  A.~Pineda, A.~Vairo,
  %``The QCD potential at O (1 / $m^{2)}$ : Complete spin dependent and spin independent result,''
  Phys.\ Rev.\  {\bf D63 } (2001)  054007.
  [hep-ph/0009145].

  
\end{thebibliography}
\end{document}